\documentclass[12pt,letterpaper]{article}

\usepackage{graphicx}
\usepackage{geometry}
\usepackage{amsmath,amsfonts,amssymb,bm}
\usepackage[hidelinks, colorlinks=false]{hyperref}
\usepackage{endnotes}
\usepackage{a4wide}
\usepackage{caption}
\usepackage{url}
\usepackage[utf8,latin1]{inputenc}
\usepackage{natbib}
\usepackage{color}
\usepackage{float}
\usepackage[nopostdot,nonumberlist,acronym]{glossaries}
\usepackage{cleveref}
\usepackage{siunitx}
\usepackage[inline]{enumitem}
\usepackage{tipa}
\usepackage{lineno}
\usepackage{lastpage}
\usepackage{fancyhdr}
    
\newacronym{hrtf}{HRTF}{head-related transfer function}
\newacronym{itd}{ITD}{interaural time difference}
\newacronym{ild}{ILD}{interaural level difference}
\newacronym{snr}{SNR}{signal-to-noise ratio}
\newacronym{srt}{SRT}{speech reception threshold}
\newacronym{ci}{CI}{cochlear implant}
\newacronym{bte}{BTE}{behind-the-ear}
\newacronym{rms}{RMS}{root-mean-square}
\newacronym{std}{s.d.}{standard deviation}
\makeglossaries

\title{Head shadow enhancement with low-frequency beamforming improves sound localization and speech perception for simulated bimodal listeners}
\author{Benjamin Dieudonn\'e*, Tom Francart}

\begin{document}
\makeatletter
\noindent
\begin{minipage}{\textwidth}
\begin{flushleft}
\Large 
\@title \\[2ex]
\large 
\@author \\[1ex]

\footnotesize
KU Leuven -- University of Leuven, Department of Neurosciences, Experimental Oto-rhino-laryngology, Herestraat 49 bus 721, B-3000 Leuven, Belgium. \\[3ex]

\normalsize
\href{mailto:benjamin.dieudonne@med.kuleuven.be}{benjamin.dieudonne@med.kuleuven.be}, \href{mailto:tom.francart@med.kuleuven.be}{tom.francart@med.kuleuven.be} \\[3ex]
Manuscript accepted by Hearing Research (March 6, 2018):
\\[0.5ex]
\url{https://doi.org/10.1016/j.heares.2018.03.007}
\end{flushleft}
\end{minipage}
\makeatother

\begin{abstract}
Many hearing-impaired listeners struggle to localize sounds due to poor availability of binaural cues. Listeners with a cochlear implant and a contralateral hearing aid -- so-called bimodal listeners -- are amongst the worst performers, as both interaural time and level differences are poorly transmitted.
We present a new method to enhance head shadow in the low frequencies. Head shadow enhancement is achieved with a fixed beamformer with contralateral attenuation in each ear. The method results in interaural level differences which vary monotonically with angle. It also improves low-frequency signal-to-noise ratios in conditions with spatially separated speech and noise.
We validated the method in two experiments with acoustic simulations of bimodal listening.
In the localization experiment, performance improved from $50.5$\textdegree\ to $26.8$\textdegree\ root-mean-square error compared with standard omni-directional microphones. 
In the speech-in-noise experiment, speech was presented from the frontal direction. Speech reception thresholds improved by $15.7$~dB~SNR when the noise was presented from the cochlear implant side, improved by $7.6$~dB~SNR when the noise was presented from the hearing aid side, and was not affected when noise was presented from all directions.
Apart from bimodal listeners, the method might also be promising for bilateral cochlear implant or hearing aid users. Its low computational complexity makes the method suitable for application in current clinical devices. \\[4ex]
\end{abstract}
\vskip 1cm
\begin{minipage}{\textwidth}
\hrulefill
\vskip 0.2cm
\small
*Corresponding author
\vskip 0.2cm

\emph{\textcopyright\ 2018. This manuscript version is made available under the CC-BY-NC-ND 4.0 license \\ \url{http://creativecommons.org/licenses/by-nc-nd/4.0/}}
\end{minipage}

\newpage
\section{Introduction}
Poor perception of binaural cues is a problem for many hearing-impaired listeners, leading to poor sound localization and speech understanding in noisy environments. Listeners with a \gls{ci} and a hearing aid in the non-implanted ear -- so-called bimodal \gls{ci} listeners -- are amongst the worst performers, as both \glspl{itd} and \glspl{ild} are poorly transmitted \citep{francart2013psychophysics}. 
\Glspl{itd} are most probably not perceived due to 
\begin{enumerate*}[label=(\arabic*)] 
\item the signal processing in clinical \gls{ci} sound processors which neglects most temporal fine structure information,
\item tonotopic mismatch between electric and acoustic stimulation, and
\item differences between the processing delay of both devices
\end{enumerate*} \citep{francart2009sensitivity,francart2011sensitivity}. 
\Glspl{ild} are also poorly perceived because 
\begin{enumerate*}[label=(\arabic*)] 
\item the head shadow is most effective at high frequencies, which are often not perceived in the non-implanted ear due to high-frequency hearing loss,
\item different dynamic range compression algorithms in both devices, and
\item different loudness growth functions for electric and acoustic stimulation
\end{enumerate*} \citep{francart2009amplification,francart2011enhancement}.
Moreover, for large angles, the natural \gls{ild}-versus-angle function becomes non-monotonic \citep{shaw1974transformation}. This means that it is physically impossible to localize sounds unambiguously for all directions with only natural \glspl{ild}.

Therefore, several authors have presented sound processing strategies to artificially enhance \glspl{ild}, resulting in improved sound localization and improved speech intelligibility in noise. 
\cite{francart2009amplification} have shown improved sound localization in an acoustic simulation of bimodal \gls{ci} listeners, by adapting the level in the hearing aid to obtain the same broadband \gls{ild} as a normal-hearing listener \citep{francart2009amplification,francart2013localisation}.
\cite{lopez2016binaural} implemented a strategy for bilateral \gls{ci} users, inspired by the contralateral medial olivocochlear reflex. Their strategy attenuated sounds in frequency regions with larger amplitude on the contralateral side, resulting in an increase in speech intelligibility for spatially separated speech and noise. 
Since both strategies are solely based on level cues that are present in the acoustic signal, they cannot solve the problem of the non-monotonic \gls{ild}-versus-angle function. 
\cite{francart2011enhancement} adapted their above-mentioned strategy by applying an artificial \gls{ild} based on the angle of incidence, to obtain a monotonic \gls{ild}-versus-angle function. They found improved sound localization for real bimodal listeners. However, this strategy relied on a priori information about the angle of incidence of the incoming sound.
\cite{brown2014binaural} extended the strategy by estimating the angle of incidence in different frequency regions based on \glspl{itd}, resulting in an improved speech intelligibility for bilateral \gls{ci} users. 
\cite{moore2016evaluation} evaluated a similar algorithm for bilateral hearing aid users, and found improved sound localization while speech perception was not significantly affected.

All above-mentioned strategies try to artificially impose an \gls{ild} based on estimations of auditory cues that are already present. Unfortunately, these estimations are either suboptimal (if based on non-monotonic \gls{ild} cues) or computationally expensive (if based on \glspl{itd}). Moreover, they can only handle multiple sound sources if these sources are temporally or spectro-temporally separated, while the spectrograms of multiple concurrent speakers most likely have some overlaps.
Recently, \cite{veugen2017effect} tried to improve the access to high-frequency \glspl{ild} for bimodal listeners without the need for estimations of auditory cues, by applying frequency compression in the hearing aid. However, they did not find a significant improvement in sound localization. Moreover, frequency compression might result in undesired side-effects on speech intelligibility, sound quality, envelope \glspl{itd} and interaural coherence \citep{simpson2009frequency,brown2016time}.

In this paper, we present and validate a novel method to enhance low-frequency \glspl{ild} without the need of estimations of auditory cues or distorting the incoming sound. We enhance the head shadow by supplying each ear with a fixed bilateral electronic beamformer applied in the low frequencies, attenuating sounds coming from its contralateral side (as opposed to conventional fixed (unilateral or bilateral) beamformers that attenuate sounds coming from the rear side). This results in enhanced low-frequency \glspl{ild} and resolves non-monotonic \gls{ild}-versus-angle functions. Because of its low computational complexity, our method is suitable for application in current clinical devices. As a proof-of-concept, we validate the effect of head shadow enhancement on localization and speech perception in noise for simulated bimodal listeners.

\section{General methods}
\subsection{Head shadow enhancement}
In the low frequencies (below $1500$~Hz), the ear naturally has an omni-directional directivity pattern, resulting in very small \glspl{ild} \citep[Chapter 7]{moore2012introduction}. We enhanced this directivity pattern with an end-fire delay-and-subtract directional microphone applied below $1500$~Hz. In each ear, the beamformer attenuated sounds arriving from its contralateral side. Above $1500$~Hz, we did not apply any beamforming. 

\begin{figure}[H]
	\centering
	\includegraphics[width=0.9\linewidth]{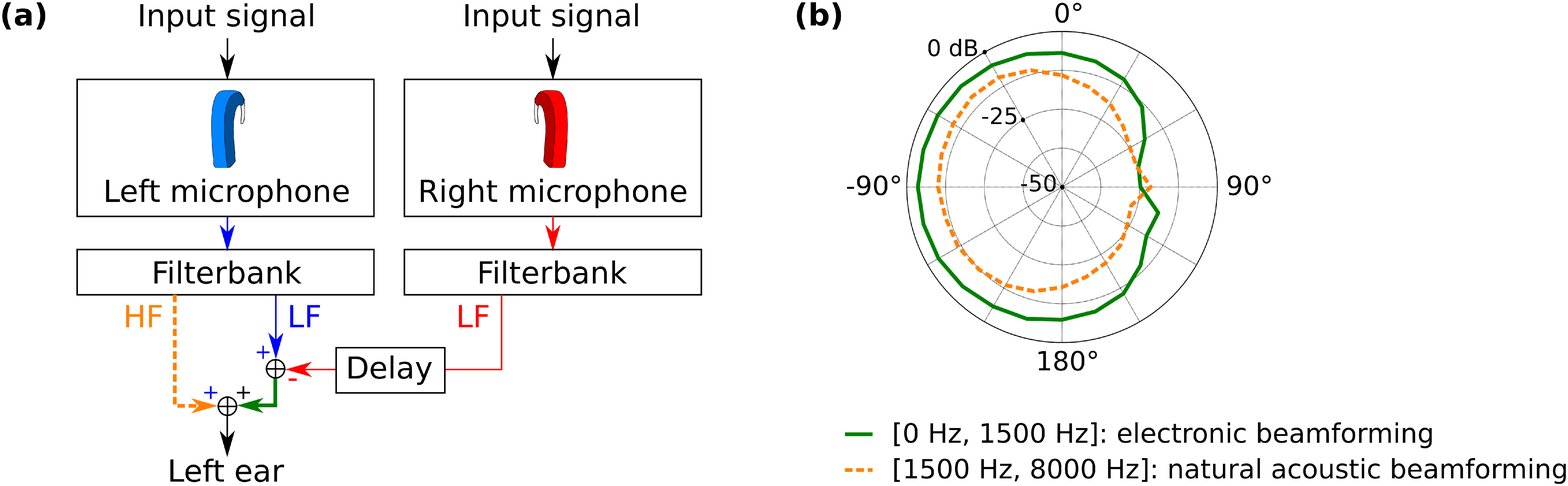}
	\caption{
	\textbf{(a)} Block diagram of head shadow enhancement algorithm. Low frequencies of the right ear signal are sent to the left ear device, followed by delay-and-subtract to obtain low-frequency contralateral attenuation. The same method is applied in the right ear device (not shown in the figure). \\
	\textbf{(b)} The method results in a cardioid-like directivity pattern for low frequencies (instead of the natural omni-directional one), while the natural directivity pattern of the ear remains unchanged for frequencies above $1500$~Hz.}
	\label{fig:algorithm}
\end{figure}

To achieve contralateral attenuation in each ear, a linear microphone array in the lateral direction was realized with a data link between the left- and right-ear devices, as illustrated in Fig.~\ref{fig:algorithm}(a). 
The low-frequency gain was boosted with a first-order low-pass Butterworth filter (cutoff at $50$~Hz), to compensate for the $6$~dB/octave attenuation of the subtractive directional microphone \citep[Chapter 7]{dillon2001hearing}.

In this set-up, the microphone spacing equals the distance between the ears, approximately $20$~cm. Such a large microphone spacing yields good sensitivity of the directional microphone in low frequencies (note that a frontal directional microphone in a \gls{bte} device is usually not active in low frequencies  because of its strong high-pass characteristic \citep{ricketts2002low}). On the other hand, this large spacing decreases the sensitivity at frequencies above approximately $800$~Hz due to the comb filter behavior of a subtractive array \citep[Chapter 7]{dillon2001hearing}: the first null in the comb filter would appear at $850$~Hz when considering a microphone distance of $20$~cm and a sound speed of $340$~m/s, the second null at $1700$~Hz, etc. This comb filtering behavior also affects the directional pattern of the beamformer. Since we only enhanced the head shadow for frequencies below $1500$~Hz, the directional pattern and frequency response were not strongly affected by the comb filtering.

In Fig.~\ref{fig:algorithm}(b) it can be seen that the method results in a cardioid-like directivity pattern for low frequencies, while the natural directivity pattern of the ear remains unchanged for frequencies above $1500$~Hz. The directivity patterns are calculated as the spectral power in the respective band with a white noise signal as input to the algorithm.

\subsection{Simulations of spatial hearing}
Spatial hearing was simulated with \glspl{hrtf}. We measured the response of an omni-directional microphone in a \gls{bte} piece placed on the right ear of a CORTEX~MK2 human-like acoustical manikin; for each angle, the left-ear \gls{hrtf} was obtained by taking the \gls{hrtf} from the right ear for a sound coming from the opposite side of the head (e.g., the left-ear \gls{hrtf} for a sound coming from $-60$\textdegree\ equaled the right-ear \gls{hrtf} for a sound coming from $+60$\textdegree). 
The manikin was positioned in the center of a localization arc with radius of approximately $1$~m, with $13$~loudspeakers (type Fostex 6301B) positioned at angles between $-90$\textdegree\ (left) and $+90$\textdegree\ (right) in steps of $15$\textdegree. To also obtain \glspl{hrtf} for sounds arriving from behind the head, we performed a second measurement in which the manikin was rotated $180$\textdegree. To simulate an anechoic response, reflections were removed by truncating each \gls{hrtf} after $2$~ms starting from its highest peak.

\subsection{Simulation of bimodal cochlear implant hearing}
We simulated bimodal \gls{ci} hearing according to the methods of \cite{francart2009amplification}. 

\gls{ci} listening was simulated in the left ear with a noise band vocoder to mimic the behavior of a \gls{ci} processor: the input signal was sent through a filter bank; within each channel, the envelope was detected with half-wave rectification followed by a $50$~Hz low-pass filter; this envelope was used to modulate a noise band of which the spectrum corresponded to the respective filter; the outputs of all channels were summed to obtain a single acoustic signal. In the localization experiment (Experiment $1$), the vocoder contained $8$ channels, logarithmically spaced between $125$~Hz and $8000$~Hz. In the speech perception experiment (Experiment $2$), we lowered the number of channels to $5$ (also logarithmically spaced between $125$~Hz and $8000$~Hz) to obtain worse speech perception, i.e., to better correspond with real \gls{ci} listening. The number of channels did not have an influence on the head shadow enhancement algorithm, as both vocoders had the same effect on the long-term spectrum of any input signal.

Severe hearing loss was simulated in the right ear with a sixth order low-pass Butterworth filter with a cutoff frequency of $500$~Hz, such that the response rolled off at $-36$~dB per octave. This corresponds with a ski-slope audiogram of a typical bimodal \gls{ci} listener. 

In this simulation with a vocoder in one ear and a low-pass filter in the other ear, little to no \gls{itd} cues could be used to localize sounds, as the vocoder removed all temporal fine structure. Note that we also ramped the on- and offset of our localization stimulus (see Section \ref{sec:stimuli}) to further reduce potential \gls{itd} cues. Therefore, our participants relied (almost) solely on \gls{ild} cues during the localization experiment \citep{francart2013psychophysics}.

\subsection{Participants}
We recruited $8$ normal-hearing participants, aged between $24$- and $26$-years-old. Their pure tone thresholds were better than or equal to $20$~dBHL at all octave frequencies between $125$ and $8000$~Hz. The study was approved by the Medical Ethical Committee of the University Hospital Leuven (S58970).

\section{Experiment 1: Localization}
\subsection{Experimental set-up}
The participant was seated in the same localization room as where the \glspl{hrtf} were measured. The loudspeakers were labeled with numbers $1$--$13$, corresponding to angles between $-90$\textdegree\ (left) and $+90$\textdegree\ (right) in steps of $15$\textdegree. The loudspeakers served solely as a visual cue. The stimuli were presented through Sennheiser HDA200 over-the-ear headphones via an RME Hammerfall DSP Multiface soundcard, using the software platform APEX~3 \citep{francart2008apex}.

\subsection{Stimuli}
\label{sec:stimuli}
A speech signal was used because of its relevance in realistic listening conditions. We presented the Dutch word ``zoem'' \textipa{[\textprimstress zum]} from the Lilliput speech material \citep{van2013lilliput}, uttered by a female talker. To limit the potential use of on-/offset \gls{itd} cues, we ramped the on- and offset with a $50$~ms cosine window. 

\subsection{Procedure}
Localization performance was measured in a condition with head shadow enhancement and a condition without head shadow enhancement; the order of conditions was randomized across subjects. Each condition consisted of a block of $7$~runs. The first $4$~runs served as training to get used to the simulation; only the last $3$~runs were considered in our analysis. Each run consisted of $3$~trials per angle, resulting in $39$~trials in total per run; the order of trials was randomized in each run. The participant was instructed to look straight ahead during stimulus presentation, and say the number indicated on the loudspeaker of the apparent sound source location after stimulus presentation. Feedback was always given after the response by turning on a light emitting diode above the correct speaker for $2$~s. Note that we did not ask the participants whether they perceived the sound image outside or inside their head.

For calibration, a speech-weighted noise with the same long-term average speech spectrum as the stimulus was constructed. The stimulus was calibrated separately for each condition, such that a signal from the front ($0$\textdegree) was presented at $65$~dBA in each ear (calibrated with a B\&K Artificial Ear Type 4153). 
To avoid the use of monaural level cues for localization, the overall level was randomly roved by $\pm10$~dB.

We used three measures to quantify localization performance (all expressed in degrees [\textdegree]): the response bias, the response \gls{std} and the \gls{rms} error. They are respectively defined as (at a certain angle for a certain subject): 

\begin{align}
\label{eq:bias}
\text{bias} & \triangleq
\left| \text{mean response} - \text{target response} \right|
\\
\label{eq:std}
\text{\gls{std}} & \triangleq
\sqrt{
\sum_{trial=1}^{\text{N}_\text{trials}}
\frac{
\left( 
\text{response}_\text{trial} - \text{mean response}
\right)^2
}{\text{N}_\text{trials}-1}
}
\\
\label{eq:rmserror}
\text{\gls{rms} error}  & \triangleq
\sqrt{
\sum_{trial=1}^{\text{N}_\text{trials}}
\frac{
\left( 
\text{response}_\text{trial} - \text{target response}
\right)^2
}{\text{N}_\text{trials}}
}
\end{align}

Both the bias and \gls{std} contribute to the \gls{rms} error. With equations \ref{eq:bias}, \ref{eq:std} and \ref{eq:rmserror}, the following equality can be deduced: 

\begin{align}
\label{eq:relation}
\text{\gls{rms} error} & =
\sqrt{
\frac{\text{N}_\text{trials}-1}{\text{N}_\text{trials}}
\text{\gls{std}}^2 +
\text{bias}^2
}
\end{align}

\subsection{Results}
\begin{figure}[H]
	\centering
	\includegraphics[width=0.9\linewidth]{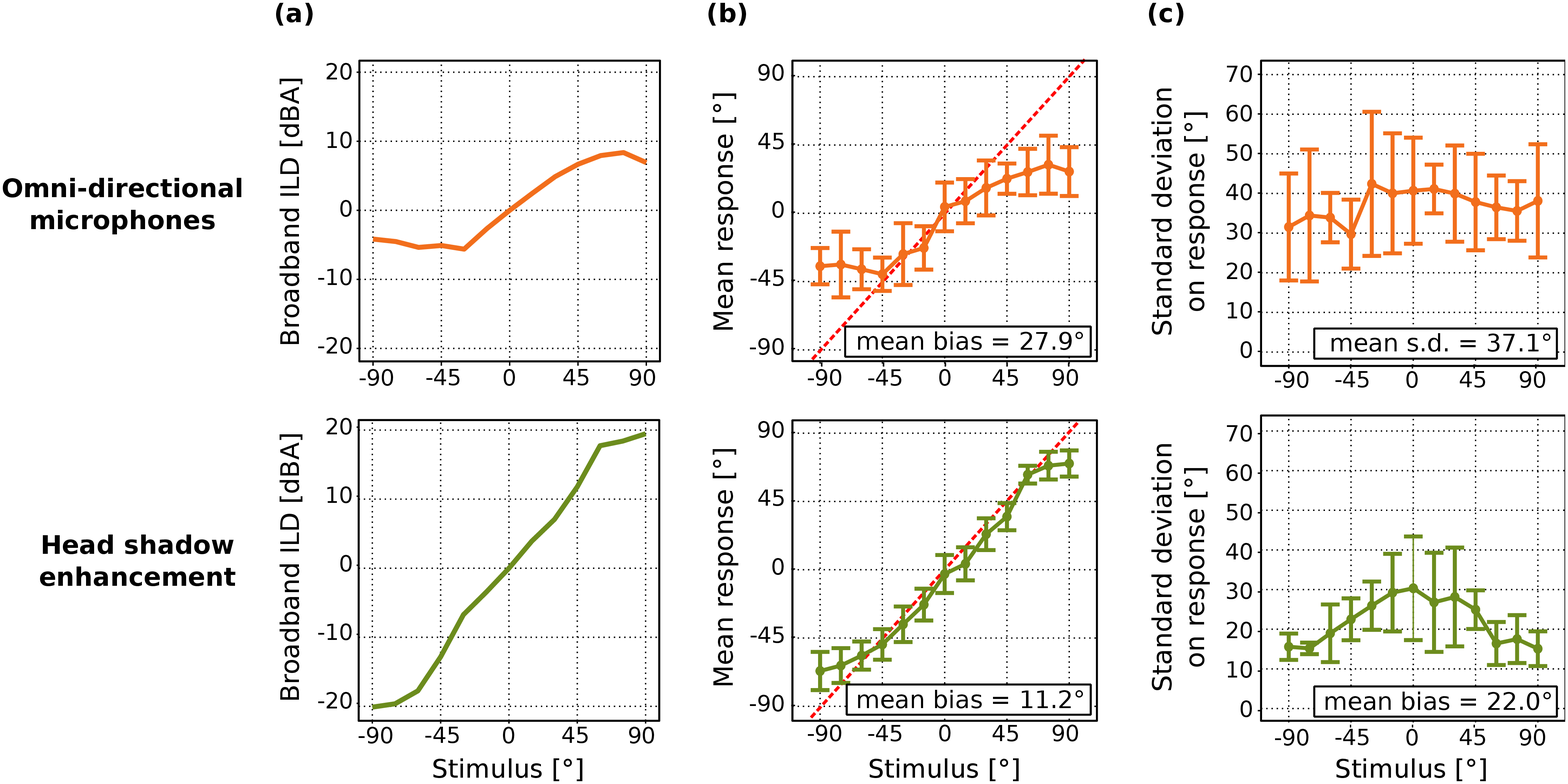}
	\caption{
Due to enhanced \acrfullpl{ild}, head shadow enhancement significantly improved localization performance by $23.7$\textdegree\ in \gls{rms} error. The \gls{rms} error is dependent on both the bias and the uncertainty in the responses. \\
\textbf{(a)} Head shadow enhancement resulted in a steeper and monotonic \gls{ild}-versus-angle curve for the speech stimulus ``zoem'' in a bimodal simulation. \\
\textbf{(b)} The mean response (averaged across trials) is a measure of the bias in the response: the closer to the diagonal (dashed line), the better the response. Head shadow enhancement decreased the bias especially for large angles, as can be expected from the \acrfull{ild} curves for our stimuli. Error bars represent the inter-subject standard deviation. \\
\textbf{(c)} The \acrfull{std} in response (across trials) is a measure of how certain the listener is of his or her response: less uncertainty results in a smaller \gls{std} Head shadow enhancement decreased the uncertainty for all angles, but especially for large angles. Error bars represent the inter-subject standard deviation.}
	\label{fig:resultslok}
\end{figure}

The broadband \glspl{ild} of the stimulus after bimodal simulation for angles between $-90$\textdegree\ and $+90$\textdegree\ with or without head shadow enhancement are shown in Fig.~\ref{fig:resultslok}(a). Head shadow enhancement resulted in a steeper and monotonic \gls{ild}-versus-angle function.

The results of the localization experiment are shown in Fig.~\ref{fig:resultslok}(b) and (c). Error bars represent the standard deviation across subjects.

In Fig.~\ref{fig:resultslok}(b), the mean response averaged across trials as a function the presentation angle is plotted per condition. This is a representation of the response bias for a certain condition: the closer the mean is to the diagonal, the smaller the response bias. 

In Fig.~\ref{fig:resultslok}(c), the response \gls{std} across trials as a function the presentation angle is plotted per condition. The response \gls{std} is a measure of the variability in the response for a certain condition: the lower the \gls{std}, the smaller the variability in the response, and thus the more certain the participants were about their response. 

It can be seen that head shadow enhancement reduces both the bias and variability in response. Both for the bias and variability, the largest improvement is for large angles, corresponding well with the \gls{ild} curves of Fig.~\ref{fig:resultslok}(a).

A Wilcoxon signed-rank test was performed to compare the \gls{rms} error averaged across all angles with or without head shadow enhancement. Head shadow enhancement significantly improved localization performance from a mean \gls{rms} error of $50.5$\textdegree\ to a mean \gls{rms} error of $26.8$\textdegree, i.e., a mean improvement of $23.7$\textdegree\ in \gls{rms} error ($\text{V}=36$, $\text{p}=0.008$, $\text{r}=-0.67$).

\subsection{Discussion}
Head shadow enhancement yielded a steeper and monotonic \gls{ild}-versus-angle function, resulting in a large improvement in sound localization of $23.7$\textdegree\ in \gls{rms} error. The strong resemblance of the \gls{ild}-versus-angle function (Figure \ref{fig:resultslok}(a)) and the mean response curve (Figure \ref{fig:resultslok}(b)) confirms that our participants were indeed relying on \glspl{ild} for localization and did not use \glspl{itd}.
The localization error (consisting of both the bias and uncertainty in response) was reduced especially at large angles, as could be expected from the \gls{ild} curves of our stimuli. Note that the unusually small \gls{std} in response at large angles (see Figure~\ref{fig:resultslok}(c)) can partly be contributed to our experimental set-up: the opportunity for erroneous responses is approximately halved at eccentric angles because there were no speakers beyond $\pm90$\textdegree.

In the condition without head shadow enhancement, we found a mean \gls{rms} error of $50.5$\textdegree, which is within the range reported for real bimodal listeners \citep{potts2009recognition,ching2007binaural}. This confirms the validity of our acoustic simulation of bimodal listening. \cite{francart2011enhancement} have indeed shown that their results for acoustic simulations of bimodal listening \citep{francart2009amplification} could be translated to real bimodal listeners.

We also expect improvements in localization for different populations, as similar methods have already been shown to be effective for real bimodal listeners \citep{francart2011enhancement} and bilateral hearing aid users \citep{moore2016evaluation}. \cite{moore2016evaluation} have even shown improvements in localization when low-frequency \gls{ild} enhancement was combined with compressive gain. Note that head shadow enhancement might distort low-frequency \glspl{itd}, which has to be taken into account when considering bilateral hearing aid users.

\section{Experiment 2: Speech perception in noise}
\subsection{Experimental set-up}
The participant was seated in a quiet room. The stimuli were again presented through Sennheiser HDA200 over-the-ear headphones via an RME Hammerfall DSP Multiface soundcard, using the software platform APEX~3 \citep{francart2008apex}.

\subsection{Stimuli}
We used the Flemish (Dutch) Matrix sentence test as target speech \citep{Luts2015}. It consists of $13$ lists of $20$ sentences uttered by a female speaker. Each sentence has the same grammatical structure (name, verb, numeral, adjective, object). As masking noise, we used stationary speech-weighted noise with the same long-term average speech spectrum as the sentences.

We measured speech perception in three spatial conditions, always with target speech from the front: noise at the \gls{ci} side (S0NCI), noise at the hearing aid side (S0NHA) and uncorrelated noise from all directions (S0N360).

\subsection{Procedure}
For each condition, we measured the \gls{srt}, defined as the \gls{snr} (at the center of the participant's head, if the experiment were done in free-field) at which $50$\% of speech could be understood. We did this according to the adaptive procedure as described by \cite{brand2002efficient}. 
The speech was presented at a level of $58$~dB~SPL during each run (calibrated with stationary speech-weighted noise with a B\&K Artificial Ear Type 4153), while the noise level was set according to the presented \gls{snr}. For each measurement, we estimated the \gls{srt} as the \gls{snr} that was calculated based on the response on the last trial. 

For each participant, we performed each measurement twice to reduce random variability in the results; before the analysis, we averaged these two repetitions for each measurement. We ended up with a total of $2$ (directional processing types) $\times$ $3$ (spatial conditions) $\times$ $2$ (repetitions) $= 12$ measurements for each subject. We performed the tests in blocks per noise direction, while randomizing the order of these blocks and randomizing the conditions within each block. Each participant started with some training lists (S0N360 without head shadow enhancement) to get used to the procedure and the bimodal simulation. 

\subsection{Results}
\begin{figure}[H]
	\centering
	\includegraphics[width=0.9\linewidth]{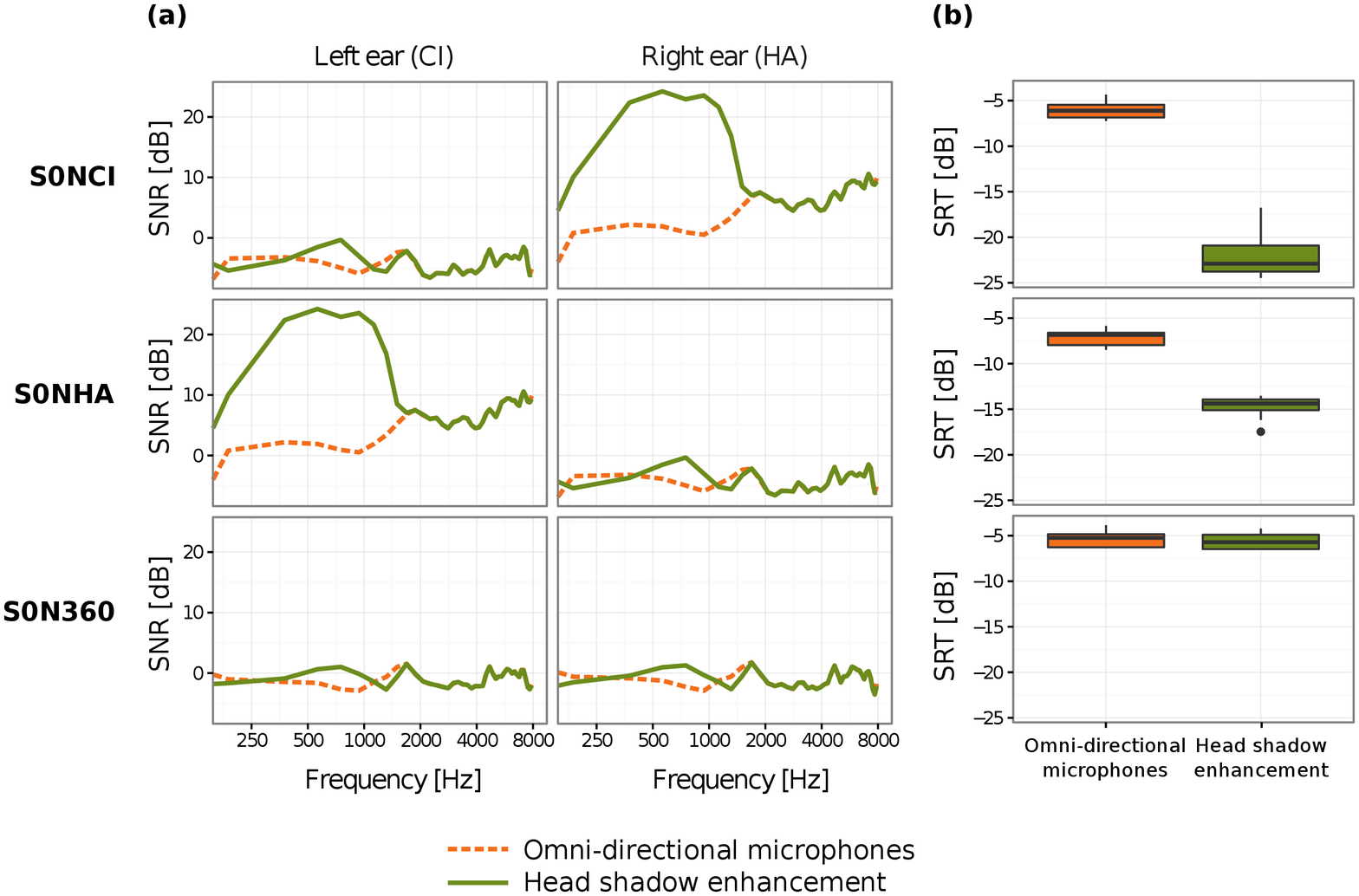}
	\caption{
	Head shadow enhancement improved \acrfull{snr} and speech understanding in spatial situations where a head shadow benefit is expected, while it never deteriorated \gls{snr} or speech intelligibility. (Note the symmetry of the \acrfullpl{hrtf} in the \gls{snr} plots.)\\
	\textbf{(a)} When the noise came from $+90$\textdegree or $-90$\textdegree (S0NHA or S0NCI), head shadow enhancement increased the \gls{snr} by up to $20$~dB~SNR at some frequencies in the ear with better \gls{snr}, while the \gls{snr} in the other ear was only slightly affected. When the noise came from all directions (S0N360), there was little to no \gls{snr}-change at either ear.\\
	\textbf{(b)} Head shadow enhancement yielded an extra head shadow benefit of $7.6$~dB~\gls{snr} (S0NHA) up to $15.7$~dB~\gls{snr} (S0NCI), while it did not affect speech intelligibility With noise from all directions (S0N360)}
	\label{fig:resultsspin}
\end{figure}

The frequency-dependent \glspl{snr} in the left and right ear for the three different spatial conditions with or without head shadow enhancement are shown in Fig.~\ref{fig:resultsspin}(a). The corresponding \glspl{srt} are shown in Fig.~\ref{fig:resultsspin}(b). For each spatial condition, we performed a Wilcoxon signed-rank test to compare the \glspl{srt} with or without head shadow enhancement.

With noise from the \gls{ci} side (S0NCI), the frequency-dependent \gls{snr} increased by up to $20$~dB in the low frequencies at the hearing aid side, while there was little to no \gls{snr}-change at the \gls{ci} side. In other words, the head shadow benefit was enhanced in the low frequencies. This resulted in a significant improvement in \gls{srt} from $-6.0$~dB~\gls{snr} to $-21.7$~dB~\gls{snr}, i.e., a mean improvement of $15.7$~dB~\gls{snr} in \gls{srt} ($\text{V}=36$, $\text{p}=0.008$, $\text{r}=-0.67$).

With noise from the hearing aid side (S0NHA), the frequency-dependent \gls{snr} increased by up to $20$~dB in the low frequencies at the \gls{ci} side, while there was little to no \gls{snr}-change at the hearing aid side. Thus, the head shadow benefit was again enhanced in the low frequencies. This resulted in a significant improvement in \gls{srt} from $-7.2$~dB~\gls{snr} to $-14.8$~dB~\gls{snr}, i.e., a mean improvement of $7.6$~dB~\gls{snr} in \gls{srt} ($\text{V}=36$, $\text{p}=0.008$, $\text{r}=-0.67$).

With noise from all directions (S0N360), there was little to no \gls{snr}-change at both ears. Consequently, there was no significant difference in \gls{srt} without or with head shadow enhancement (\glspl{srt} were $-5.3$ and $-5.6$~dB~SNR respectively, $\text{V}=36$, $\text{p}=0.38$, $\text{r}=-0.22$).

\subsection{Discussion}
In spatial situations where one would expect head shadow benefits (S0NCI and S0NHA), head shadow enhancement yielded a large increase in \gls{snr} in the ear with better \gls{snr}, while there was little to no effect on the \gls{snr} in the other ear. This resulted in a large improvement in speech intelligibility: an extra head shadow benefit of $7.6$~dB~\gls{snr} (S0NHA) up to $15.7$~dB~\gls{snr} (S0NCI). Thus, the largest benefit was obtained when the hearing aid side was the ear with the better \gls{snr} (S0NCI). On the one hand, this makes sense, as the algorithm works for the whole hearing spectrum at this side ($0$ to $500$~Hz). On the other hand, this implies that our participants were relying mostly on the hearing aid side to understand speech in S0NCI, which might not correspond with a typical bimodal listener with profound hearing loss in the non-implanted ear. While the frequency range of $0$ to $500$~Hz does correspond quite well with the residual hearing of a typical bimodal listener, we did not take into account degraded frequency selectivity, cognitive ability, etc. Moreover, we used a closed-set speech material \citep[Flemish Matrix,][]{Luts2015}, which might have facilitated speech understanding with this narrow hearing spectrum.

With noise from all directions (S0N360), there was little to no effect on the \gls{snr} in both ears. Although the beamformer yields a small frontal attenuation, it also attenuates noise from the contralateral side in each ear. Consequently, there was no net \gls{snr} change with or without head shadow enhancement, neither a significant difference in speech intelligibility.

In the conditions without head shadow enhancement, we found \glspl{srt} around $-6$~dB, which corresponds to the best performers in a real bimodal population \citep{devocht2017benefits} (note that \cite{devocht2017benefits} used the Dutch Matrix sentence test and not the Flemish one). This again confirms the validity of our acoustic simulation of bimodal listening. 

\section{General discussion}
The current study shows the possible effectiveness of a head shadow enhancement algorithm based on fixed beamformers. The algorithm is able to enhance \glspl{ild} and \glspl{snr} by supplying each ear with a fixed beamformer with contralateral attenuation. We found large improvements in localization ability and speech understanding for simulated bimodal listeners.

We believe that our results can be translated to real bimodal listeners, as performance in our acoustic simulations without head shadow enhancement corresponded well with the performance of real bimodal listeners. Note however that acoustic simulations with young normal-hearing listeners do not take into account all aspects of real bimodal listening, such as degraded frequency selectivity in the ear with residual hearing and cognitive abilities. Moreover, the population of real bimodal listeners has a large variability in performance. Future investigations should determine how their baseline performance interacts with the benefit of head shadow enhancement. The method might also be detrimental for some part of the population, such as listeners with extremely poor hearing in one ear: if target speech is then presented to the non-implanted ear, they will not be able to rely on their implanted ear because of the strong contralateral attenuation.

The method is also promising for other device configurations, as similar approaches have been shown effective in improving localization performance and speech intelligibility for bilateral hearing aid users \citep{moore2016evaluation} and bilateral cochlear implant users \citep{lopez2016binaural,brown2014binaural}. The benefit might even be larger than expected, as improved localization allows listeners to orient towards talkers and gain access to visual cues, resulting in an additional improvement in speech intelligibility \citep{van2015audio}. Our method is distinguished from previously reported strategies due to its simplicity and low computational complexity. The latter makes it also suitable for application in clinical devices. 

Future investigations should ensure how the algorithm interacts with different acoustics and sound processing blocks:
\begin{enumerate}[label=\arabic*.] 
\item We expect that the algorithm will have no difficulties with multiple sound sources, as it is based on a fixed beamformer which naturally handles multiple sources. 
\item The effect on \glspl{itd} remains to be investigated. As both bilateral and bimodal \gls{ci} users hardly perceive \glspl{itd}, any detrimental effect on \glspl{itd} should not be an issue for this population. However, it might decrease performance for bilateral hearing aid users.
\item The effect of low-frequency \glspl{ild} for very close sound sources \citep[closer than $1$~m,][]{brungart1999auditory} also remains to be investigated; however, they may be dealt with by a time-dependent comparator that equalizes left and right microphone signals before the delay-and-subtract takes place.
\item The benefit of head shadow enhancement will most probably depend on the spectrum of the sources: the more information that is carried in the low frequencies, the larger the effect of the beamformer. We expect listeners to be able to adapt to altered \gls{ild} cues for different source spectra \citep{francart2011enhancement}.
\item The beamformer can be combined with any other (monaural) signal processing block, as long as the processing does not strongly distort the signal and the total processing delay is the same in both ears. For optimal performance, it is probably recommended to have head shadow enhancement as a first block in the processing chain. Only frontal directivity has to be applied before head shadow enhancement, as otherwise head shadow enhancement should be applied to both front and rear microphone signals before applying frontal directivity. This would require double the amount of (wireless) data transfer between the two devices, and reduce battery life. Note that frontal directivity is mostly active in higher frequencies, which reduces even more the possibility of any deteriorating interaction with head shadow enhancement.
\item Combination of head shadow enhancement with binaural beamformers with frontal directivity is not as straightforward, as those beamformers often use the $4$ available microphones to end up with $1$ signal that is presented diotically \citep{buechner2014advanced}. It should however be possible to combine head shadow enhancement with a binaural beamformer that preserves (enhanced) binaural cues. Those designs typically trade off between noise reduction and binaural cue preservation \citep{van2009speech}.
\end{enumerate} 

\section{Conclusions}
We presented a new method to enhance head shadow in low frequencies, with a fixed beamformer with contralateral attenuation in each ear. 
Head shadow enhancement improved localization performance by almost $24\textdegree$ \gls{rms} error relative to $50\textdegree$ \gls{rms} error for simulated bimodal \gls{ci} listeners. It also improved speech intelligibility by up to $15.7$~dB~SNR in spatial conditions where head shadow is expected to be present, while it never deteriorated speech understanding.
The method is also promising for other hearing-impaired populations, such as bilateral cochlear implant users or bilateral hearing aid users.
Its low computational complexity makes it suitable for application in clinical devices. 

\section*{Acknowledgments}
This research is funded by the Research Foundation -- Flanders (SB PhD fellow at FWO); this research is jointly funded by Cochlear Ltd. and Flanders Innovation \& Entrepreneurship (formerly IWT), project 150432; this project has also received funding from the European Research Council (ERC) under the European Union's Horizon 2020 research and innovation programme (grant agreement No 637424, ERC starting Grant to Tom Francart). We thank our participants for their patience and enthusiasm during our experiment.

\end{document}